\documentclass{ifacconf}
\usepackage{graphicx}                                   
\usepackage{natbib}                                     
\usepackage{amsfonts}                                   
\usepackage{amsmath}                                    
\usepackage[short,c2]{optidef}                          
\usepackage[algo2e,ruled,linesnumbered]{algorithm2e}    
\usepackage[utf8]{inputenc}                             
\usepackage{pgfplots}
\DeclareUnicodeCharacter{2212}{−}
\usepgfplotslibrary{groupplots,dateplot}
\usetikzlibrary{patterns,shapes.arrows}
\pgfplotsset{compat=newest}

\newtheorem{assumption}{Assumption}                     
\usepackage[inline]{enumitem}

\newcommand{\norm}[2][]{{\left\lVert#2\right\rVert}_{#1}}
\newcommand{\expval}[2][]{\mathbb{E}_{#1}\left[#2\right]}
\newcommand{\probab}[2][]{\ifthenelse{\isempty{#2}}{\mathbb{P}_{#1}}{\mathbb{P}_{#1}\left[#2\right]}}

\begin{document}

\begin{frontmatter}
    \title{Learning safety in model-based Reinforcement Learning using MPC and Gaussian Processes}

    \author{Filippo Airaldi \ }
    \author{Bart De Schutter \ }
    \author{Azita Dabiri}

    \address{Delft Center for Systems and Control, Delft University of Technology, Mekelweg 2, 2628 CD Delft, The Netherlands \\
        Emails: \{f.airaldi,b.deschutter,a.dabiri\}@tudelft.nl}

    \begin{abstract} 
        We propose a method to encourage safety in Model Predictive Control (MPC)-based Reinforcement Learning (RL) via Gaussian Process (GP) regression. This framework consists of
\begin{enumerate*}[label=\arabic*)]  
    \item a parametric MPC scheme that is employed as model-based controller with approximate knowledge on the real system's dynamics,
    \item an episodic RL algorithm tasked with adjusting the MPC parametrization in order to increase its performance, and lastly,
    \item GP regressors used to estimate, directly from data, constraints on the MPC parameters capable of predicting, up to some probability, whether the parametrization is likely to yield a safe or unsafe policy.
\end{enumerate*}
These constraints are then enforced onto the RL updates in an effort to enhance the learning method with a probabilistic safety mechanism. Compared to other recent publications combining safe RL with MPC, our method does not require further assumptions on, e.g., the prediction model in order to retain computational tractability. We illustrate the results of our method in a numerical example on the control of a quadrotor drone in a safety-critical environment.

    \end{abstract}

    \begin{keyword} 
        Learning-based Model Predictive Control, Safe Reinforcement Learning, Gaussian Processes
    \end{keyword}
\end{frontmatter}

\section{Introduction} \label{section:introduction}
The modern proliferation of advances in Machine Learning, combined with the increasing availability of computational and sensing capabilities in modern control systems, has led to a growing interest in learning-based control methodologies that strive to learn different elements of the controller scheme from data in order to, e.g., reduce conservativeness and improve closed-loop performance, or encourage safety and robustness \citep{brunke_safe_2022}. In particular, Model Predictive Control (MPC) is the subject of a large amount of the current research in this field thanks to its wide variety of applications and high versatility, especially in regard to multivariate and constrained systems \citep{hewing_learning_2020}. While robust and stochastic MPC schemes allow for a systematic handling of different sources of uncertainties affecting the MPC performance, they often follow a strict separation between the offline design phase and the closed-loop application, where in most cases the controller remains unresponsive to changes in, e.g., the system or task. Adaptive and learning-based methodologies aim at overcoming this paradigm by automating the controller design and adaptation based on data collected during operation \citep{mesbah_2022_fusion}.

To this end, several tools from ML have been successfully coupled with MPC in order to enhance its performance or augment its capabilities. Among all, we can refer to Gaussian Process (GP) regression, which has been widely employed to, e.g., automatically learn the unmodelled (often nonlinear) dynamics in the prediction model \citep{hewing_cautious_2020} and to construct provably accurate confidence intervals on predicted trajectories \citep{koller_learning_2018}; and black-box Bayesian Optimization, which is a widespread tool for automatic tuning and optimization of the MPC controller hyperparameters \citep{piga_performance_2019}, thus relieving some burden on the design phase. Equally interesting is the combination of MPC with Reinforcement Learning (RL), another paradigm of ML gaining more and more attention thanks to its recent developments, in which an agent learn how to interact with the environment and take actions so as to maximize/minimize a reward/cost signal \citep{sutton_reinforcement_2018}. The majority of RL algorithms that are nowadays popular in the ML community are model-free and rely purely on observed state transitions and realizations of stage cost to increase the performance of the control policy. Deep Neural Networks (DNN) are predominantly used as function approximators and, while they have proven effective, they lack interpretability, and their behaviour is difficult to analyse, certify, and trust in regard to stability and constraint satisfaction.

Recently, \cite{gros_datadriven_2020} proposed and justified the use of MPC as the function approximation of the optimal policy in model-based RL. In such a scheme, the MPC controller yields the policy and value functions via its optimization problem, while, at the same time, the learning algorithm is tasked with adjusting the parametrization of the controller in an effort to implicitly or explicitly discover the optimal policy, thus improving closed-loop performance in a data-driven fashion. Via sensitivity analysis \citep{buskens_sensitivity_2001}, it is possible to apply RL algorithms such as Q-learning and stochastic/deterministic policy gradient by differentiating the MPC optimization problem with respect to its parameters and updating these accordingly. In contrast to DNN-based RL, MPC-based RL agents can ideally exploit the underlying optimization problem to provide stability requirements by construction and guarantee safety, which defines the ability of a control policy to yield state-action trajectories that do not violate system constraints. However, even in this novel paradigm, certificates on safety and stability remain an arduous challenge. If proper care is not taken, the RL updates can still potentially jeopardize the MPC scheme by updating its parameters to an undesirable configuration, resulting in a controller with unsatisfactory behaviour, e.g., that does violate constraints and hence is unsafe.

In this paper, we propose and explore a learning method in the context of safe MPC-based RL where the estimation of the safety set (i.e., the set of MPC parameters that yields a controller able to attain constraint satisfaction) is carried out in a data-driven fashion. In particular, we borrow a technique for estimating unknown constraints from the fields of constrained Bayesian Optimization \citep{sorourifar_data_2021} and safe exploration \citep{schreiter_safe_2015}, which offer tools to tackle black-box optimization problems that are constrained by some unknown functions in the decision variables. As a consequence, we are able to learn in a probabilistic way which configurations of the MPC controller are safe or not directly from data.

The rest of this article is organized as follows. Background and motivations both for safe RL in combination with MPC and for unknown constraint modelling via GP are given in Section \ref{section:background}. The proposed algorithm and details of its implementation are discussed in Section \ref{section:algorithm}. A numerical experiment showing the effectiveness of the proposed method in a safe-critical quadrotor control application is reported and analysed in Section \ref{section:experiment}. Lastly, conclusions and directions for future research are presented in Section \ref{section:conclusions}.

\section{Background} \label{section:background}
\subsection{MPC as Function Approximation in Safe RL}
As in \cite{busoniu_reinforcement_2017,sutton_reinforcement_2018}, we describe discrete-time system dynamics as a Markov Decision Process with continuous state $s \in \mathbb{R}^{n_s}$ and continuous action $a \in \mathbb{R}^{n_a}$, and state transitions $s,a \rightarrow s_+$ with the underlying conditional probability density
\begin{equation} \label{eq:background:rl:transition_probab}
    \probab{s_+ | s, a} : \mathbb{R}^{n_s} \times \mathbb{R}^{n_s} \times \mathbb{R}^{n_a} \rightarrow [0, 1].
\end{equation}
Considering a deterministic policy $\pi_\theta(s) : \mathbb{R}^{n_s} \rightarrow \mathbb{R}^{n_a}$ parametrized in $\theta \in \mathbb{R}^{n_\theta}$ and resulting in the state distribution $\tau_{\pi_\theta}$. The performance of such policy is defined as
\begin{equation}
    J(\pi_\theta) \coloneqq \expval[\tau_{\pi_\theta}]{\sum_{k=0}^{\infty}{\gamma^k L \bigl(s_k, \pi_\theta(s_k)\bigr)}},
\end{equation}
where $s_k$ is the state at time step $k$, $L(s,a) : \mathbb{R}^{n_s} \times \mathbb{R}^{n_a} \rightarrow \mathbb{R}$ is the stage cost and $\gamma \in (0, 1]$ is the discount factor. The RL problem is then to find the optimal policy $\pi^\star_\theta$ as
\begin{argmini}
    { \theta }{ J(\pi_\theta)}{ \label{eq:background:rl:optimal_policy} }{ \pi_\theta^\star = }.
\end{argmini}
While DNNs are possibly the most common choice for representing such policy, an MPC scheme can be exploited as function approximation as proposed in \cite{gros_datadriven_2020}. Consider the following MPC problem approximating the value function $V_\theta : \mathbb{R}^{n_s} \rightarrow \mathbb{R}$ as
\begin{mini!}
{ u, x, \sigma }{
    \lambda_\theta(x_0) +
    \sum_{k=0}^{N-1}{\gamma^k \left(L_\theta(x_k, u_k) + w^\top \sigma_k\right)}
    \notag
}{ \label{eq:background:V:scheme} }{ V_\theta(s) = }
\breakObjective{ \hspace{1.35cm} + \gamma^N \left(V_\theta^\text{f}(x_N) + w_\text{f}^\top \sigma_N\right) \label{eq:background:V:obj} }
\addConstraint{ x_{k+1} }{ = f_\theta(x_k, u_k) \label{eq:background:V:constraint:dyn} }{}
\addConstraint{ h_\theta(x_k, u_k) }{ \le \sigma_k \label{eq:background:V:constraint:h} }{}
\addConstraint{ h_\theta^\text{f}(x_N) }{ \le \sigma_N \label{eq:background:V:constraint:h_final} }{}
\addConstraint{ x_0 }{ = s \label{eq:background:V:constraint:x0} }{}
\end{mini!}
where vectors $x$, $u$ and $\sigma$ respectively collect states, actions and slack variables over the horizon $N$. In \eqref{eq:background:V:obj}, $\lambda_\theta(x_0)$ is an initial cost term, $L_\theta(x,u)$ is the stage cost, and $V_\theta^\text{f}(x)$ is a terminal cost approximation. $f_\theta(x,u)$ is the model approximation, and $h_\theta(x,u)$, $h^\text{f}_\theta(x)$ are inequality constraints. Lastly, $w$ and $w_{\text{f}}$ are the weights of the slack variable in the objective. The value function \eqref{eq:background:V:scheme} satisfies the fundamental equalities of the Bellman equations:
\begin{mini!}
{ u, x, \sigma }{ \eqref{eq:background:V:obj} }{}{ Q_\theta(s, a) = }
\addConstraint{ \eqref{eq:background:V:constraint:dyn}-\eqref{eq:background:V:constraint:x0} }{}{}
\addConstraint{ u_0 }{ = a }{},
\end{mini!}
\begin{argmini}
    { a }{ Q_\theta(s, a) }{}{ \pi_\theta(s) = }.
\end{argmini}

Therefore, in RL terms, the MPC parametric scheme acts as policy provider for the learning agent, whose goal is to modify the parameters $\theta$ of the controller in an effort to minimize \eqref{eq:background:rl:optimal_policy}. Various forms of RL exist that solve this problem directly or indirectly via iterative updates
\begin{equation} \label{eq:background:rl:update}
    \theta \leftarrow \theta - \alpha \nabla_\theta \sum_{k=0}^{m}{\psi(s_k,a_k,s_{k+1},\theta)},
\end{equation}
where $\alpha \in \mathbb{R}_+$ is the learning rate, $m$ denotes the number of observations used in the update (i.e., a batch of observations), and $\psi$ captures the controller's performance and varies with the specific algorithm. For example, in recursive Q-learning we have that $m=1$ and
\begin{equation}
    \psi(s_k,a_k,s_{k+1},\theta) = \bigl(L(s_k,a_k) + \gamma V_\theta(s_{k+1}) - Q_\theta(s_k,a_k)\bigr)^2,
\end{equation}
whereas in policy gradient methods
\begin{equation}
    \expval[\tau_{\pi_\theta}]{ \psi(s_k,a_k,s_{k+1},\theta) } = J(\pi_\theta).
\end{equation}

Nevertheless, while in most learning algorithms a safe policy is usually implicitly achieved at convergence assuming violations are appropriately penalized, naively performing updates \eqref{eq:background:rl:update} does not guarantee such property during the learning process itself. Thus, we would like to restrict the search of $\theta$ to a set $\mathcal{S}$ that ensures the safety of the corresponding parametric policy. More specifically, consider a generic episodic task of duration $T$ with $n_\text{c}$ safety-critical constraints, gathered as the function $h : \mathbb{R}^{n_s} \times \mathbb{R}^{n_a} \rightarrow \mathbb{R}^{n_\text{c}}$ (e.g., bounds on the temperature of a chemical reactor, or obstacles in autonomous navigation). By deploying the MPC policy $\pi_\theta$ to accomplish the task, safety amounts to satisfying
\begin{equation} \label{eq:background:safety_constraints}
    h\bigl(x_t, \pi_\theta(x_t)\bigr) \le 0, \ t=1,\dots,T.
\end{equation}
Thus, we can define the safe set $\mathcal{S}$ as
\begin{equation} \label{eq:background:safe_safe}
    \mathcal{S} \coloneqq \left\{ \theta \ | \ \eqref{eq:background:safety_constraints} \text{ holds} \right\},
\end{equation}
leading to the following reformulation of \eqref{eq:background:rl:update} as a constrained update
\begin{mini!}
{ \theta_+ }{
\frac{1}{2} \norm[2]{\theta_+ - \theta}^2 +
\alpha \nabla_\theta \sum_{k=0}^{m}{\psi(s_k, a_k, s_{k+1},\theta)}^\top (\theta_+ - \theta)
\label{eq:background:rl:constrained_update:obj}
}{ \label{eq:background:rl:constrained_update} }{}
\addConstraint{ \theta_+ \in \mathcal{S} }{}{},
\end{mini!}
by then setting $\theta \leftarrow \theta^\star_+$, where $\theta^\star_+$ is the optimal point of \eqref{eq:background:rl:constrained_update}. Nonetheless, the characterization of $\mathcal{S}$ is in general difficult, even more so in the case of function approximators with vast parametrizations. Fortunately, an MPC approximator can be leveraged here: as shown in \cite{zanon_safe_2021} and further analysed in terms of feasibility and stability in \cite{gros_learning_2022}, a robust MPC formulation coupled with an online set-membership system identification approach allows to guarantee that the MPC-RL algorithm is robustly safe with respect to the estimated disturbance set throughout the whole training, despite the fact that the dynamics of the real system are unknown to the learning agent. However, even though an adaptive robust scheme is deployed, safety is only guaranteed up to a probability $< 1$, due to the lack of perfect knowledge of the underlying system \eqref{eq:background:rl:transition_probab}. In this context, safety can be viewed in the light of Bayesian inference, where it is probabilistically conditioned on our limited knowledge of the system, i.e., prior information and actual data, labelled as $\mathcal{D}$, with probability $\beta$. Henceforth, we will refer to the notion of safety in a probabilistic term and define $\mathcal{S}_\mathcal{D}$ as
\begin{equation} \label{eq:background:probab_safe_safe}
    \mathcal{S}_\mathcal{D} \coloneqq \left\{ \theta \ | \ \probab{\eqref{eq:background:safety_constraints} | \mathcal{D} } \ge \beta  \right\}.
\end{equation}
A limitation in the approach proposed by \cite{zanon_safe_2021} is that, in order to retain computational tractability in the algorithm, the MPC prediction model $f_\theta$ has to be linear and kept fixed during learning. As acknowledged in the paper \citep{zanon_safe_2021}, while linear MPC applied to nonlinear systems can still be successful, strong nonlinearities are likely to deteriorate its performance, compromise safety, and prevent the agent from tuning the prediction model, which might severely hinder learning and lead to convergence to suboptimal policies.

In this paper, we propose a method that estimates the safe set $\mathcal{S}_\mathcal{D}$ directly from episodic data in a way that bypasses the aforementioned limitation and that conforms with the probabilistic nature of the issue of safety itself. As detailed next, we propose to leverage GP regression tools that deliver surrogate models that are able to provide uncertainty estimates whether a specific $\theta$ belongs to $\mathcal{S}_\mathcal{D}$ or not. This will then allow us to formulate an RL algorithm that can learn while being safe, up to some probability.

\subsection{Estimation of Unknown Constraints}
In essence, the matter of the estimation of set $\mathcal{S}_\mathcal{D}$ in order to impose safety coincides with estimating whether a given state-action trajectory, produced by an MPC controller parametrized in $\theta$, will satisfy constraints $h(x_t, u_t) \le 0, \ t=1,\dots,T$, with $u_t = \pi_\theta(x_t)$. However, while on the one hand the relationship $h(x, u)$ is given by the task, on the other hand we cannot predict beforehand whether $\pi_\theta(x)$ will yield safe control actions and hence safe trajectories. Therefore, we introduce the following safety constraint function $z : \mathbb{R}^{n_\theta} \rightarrow \mathbb{R}^{n_\text{c}}$
\begin{equation} \label{eq:background:max_safety_constraints}
    z(\theta) = \max_{t = 1,\dots,T} h\bigl(x_t, \pi_\theta(x_t)\bigr)
\end{equation}
that directly maps parametrizations to positive values, in case $\theta$ resulted in trajectories with violations, or to negative values, in case of safe trajectories. Analytically, this function is unknown to us, but each time a state-action trajectory for the given task is computed, it can be evaluated for the current $\theta$. Therefore, by leveraging past observed data, an approximation of \eqref{eq:background:max_safety_constraints} can be learned and used to predict safety at any previously unobserved query point $\theta^\ast$. Thus, we could use this surrogate model of $z(\theta)$ to constrain learning only to those parameters that are estimated to be safe.
\begin{assumption}
    We assume the trajectory to be dependent only on $\theta$, and, consequently, that $z(\theta)$ wholly captures safety (i.e., no other factor has an impact).
\end{assumption}
\begin{rem}
    In general, this is not the case as also the initial and final conditions of the task, as well as the initialization of the optimization solver in case of nonlinear MPC, will affect constraint satisfaction. To mitigate these dependencies, we assume the episodic task to be repetitive, e.g., to have the same initial and final conditions, and the MPC problem to be convex. Analogously, we could also assume these factors to be readily included into the MPC parametrization $\theta$ so that, in principle, all parameters affecting safety may be included in $z(\theta)$ and made learnable.
\end{rem}
\begin{rem}
    In \eqref{eq:background:max_safety_constraints}, the maximum over the time horizon $T$ is taken in order to reduce violations along a trajectory to a point-wise maximum violation, but this does not invalidate \eqref{eq:background:probab_safe_safe} since $z(\theta) \le 0 \Leftrightarrow h\bigl(x_t, \pi_\theta(x_t)\bigr) \le 0, \ t=1,\dots,T$. Likewise, functions other than $\max$ could be crafted to quantify safety of a trajectory while still encoding \eqref{eq:background:probab_safe_safe}.
\end{rem}

Unknown constraint estimation has been already tackled in literature. In constrained Bayesian Optimization, \cite{sorourifar_data_2021} propose a method for black-box optimization with unknown constraints that can only be evaluated at specific query points, where GP regression is exploited to achieve surrogate models of both the objective and the constraint functions, the latter being then used to weigh the exploration strategy by the probability of constraint satisfaction; conversely, \cite{krishnamoorthy_safe_2022} include the GP models as barrier penalty terms. Safe exploration algorithms also make extensive use of GPs, either for classification \citep{schreiter_safe_2015} or in combination with regularity tools \citep{turchetta_safe_2019}. While GPs are a popular choice because they naturally allow for a quantification of the prediction uncertainty, \cite{zhu_cglisp_2022} suggest modelling the unknown functions via Inverse Distance Weighting interpolation, which are employed to directly predict the probability of being feasible.

In this paper, GPs are used to model the safety-critical constraints \eqref{eq:background:max_safety_constraints} since these processes allow to inherently address the probabilistic nature of \eqref{eq:background:probab_safe_safe}. Training data for the GPs are collected each time the given task is solved, yielding the tuple $\langle \theta, \tilde{z}\rangle$, where $\theta$ is the current MPC parametrization and $\tilde{z}$ is a noisy observation of \eqref{eq:background:max_safety_constraints}. Observations are generally regarded as noisy due to, e.g., measurement noise or disturbances in the dynamics. Therefore, the basic idea is that, given the dataset $\mathcal{D}_n = \{\left(\theta^i, \tilde{z}^i\right)\}^{n}_{i=1}$, one independent GP regressor per constraint is used to model the corresponding safety constraint function $z_j, \ j=1,\dots,n_\text{c}$. In particular, we assume the function to have a GP prior with mean $\mu_0 : \mathbb{R}^{n_\theta} \rightarrow \mathbb{R}$ and covariance kernel $\nu : \mathbb{R}^{n_\theta} \times \mathbb{R}^{n_\theta} \rightarrow \mathbb{R}$. Under the GP prior and assumed i.i.d. Gaussian noise, observations $\tilde{z}^i_j$ are jointly Gaussian distributed, so that, at any query point $\theta^\ast$, the corresponding constraint violation $z_j(\theta^\ast)$ must be jointly Gaussian too, i.e., $z_j(\theta^\ast) \sim \mathcal{N}\bigl(\mu_j^n(\theta^\ast), {\sigma_j^n}^2(\theta^\ast)\bigr)$, where $\mu_j^n(\theta^\ast)$ is the estimate mean and $\sigma_j^n(\theta^\ast)$ its uncertainty \cite[Algorithm~2.1]{rasmussen_gaussian_2005}. In this context, \eqref{eq:background:probab_safe_safe} can be approximated from data as
\begin{equation}
    \mathcal{S}_{\mathcal{D}_n} = \left\{ \theta \ | \ \probab{z_j(\theta) \le 0 | \mathcal{D}_n} \ge \beta, \ j=1,\dots,n_\text{c} \right\},
\end{equation}
where $z_j$ is represented by a GP trained on data in $\mathcal{D}_n$.

In the next section, we illustrate how to combine MPC-based RL with GP constraint modelling into a safety-aware algorithm.

\section{Data-driven Safe MPC-based RL} \label{section:algorithm}
Algorithm \ref{algo:algorithm:main} illustrates the procedure to promote safety, up to some probability $\beta$, upon an MPC-based RL agent. Starting from an initial parametrization of the MPC controller and a (possibly empty) dataset of initial observations, experiment $i$ is carried out by completing the given task with an MPC scheme parametrized in $\theta^i$. At the end of such experiment, the full state-action trajectory can be collected and the maximum constraint violation $\tilde{z}^i$ computed. Thanks to the collected data, constraints $z_j(\theta), \ j = 1,\dots,n_\text{c}$ are then approximated via GP regressors. The RL update is then enhanced with the trained GP-based constraints, which help the algorithm in selecting the next parameters $\theta_+$ that satisfy these constraints and thus are likely to yield a safe policy.

{
    \setlength{\algomargin}{1.5em}
    \begin{algorithm2e}
        \label{algo:algorithm:main}
        \SetKwRepeat{Do}{do}{while}
        \SetKw{Break}{break}
        \DontPrintSemicolon
        \caption{Data-driven Safe MPC-based RL}

        \KwIn{Initial MPC parameters $\theta_0$; initial observations $\mathcal{D}_0$; $\varrho \in (0, 1)$}

        \For{$i = 1,\dots,n_{\max}$}{
            Perform MPC closed-loop task with $\theta^i$ by solving \eqref{eq:background:V:scheme} at each time step $t=1,\dots,T$

            Observe $\tilde{z}_j^i, \ j=1,\dots,n_\text{c}$ as per \eqref{eq:background:max_safety_constraints}

            Augment dataset $\mathcal{D}_i \leftarrow \mathcal{D}_{i-1} \cup \{\left(\theta^i, \tilde{z}^i\right)\}$

            Train GPs for $z_j(\theta), \ j=1,\dots,n_\text{c}$ on $\mathcal{D}_i$

            \Do{\eqref{eq:algorithm:constrainted_update} is \textbf{not} feasible \label{line:algorithm:backtracking_endline}}{ \label{line:algorithm:backtracking_startline}
                Try performing safe RL update
                \begin{argmini!}
                    { \theta_+ }{ \eqref{eq:background:rl:constrained_update:obj} }{ \label{eq:algorithm:constrainted_update} }{ \theta_+^\star \leftarrow }
                    \addConstraint{ \theta_+ \in \mathcal{S}_{\mathcal{D}_i} }{ \label{eq:algorithm:constrainted_update:gp_constraints} }{}
                \end{argmini!}

                Reduce safety probability $\beta \leftarrow \varrho \beta$
            }



            Update parametrization $\theta_{i+1} \leftarrow \theta_+^\star$
        }
    \end{algorithm2e}
}

\begin{rem} \label{remark:algorithm:backtracking_beta}
    Gaussianity of the GPs can be conveniently leveraged to express constraint \eqref{eq:algorithm:constrainted_update:gp_constraints} deterministically as
    \begin{equation}
        \mu_j^i(\theta_+) + \Phi^{-1}(\beta) \sigma_j^i(\theta_+) \le 0, \ j=1,\dots,n_\text{c},
    \end{equation}
    where $\Phi^{-1}$ is the normal distribution quantile function. However, since the GP terms $\mu_j^i(\theta_+)$ and $\sigma_j^i(\theta_+)$ are in practice always nonlinear, the overall RL update optimization problem \eqref{eq:algorithm:constrainted_update} becomes nonlinear too. Moreover, there exist no guarantees on the feasibility of \eqref{eq:algorithm:constrainted_update}, i.e., the update might get stuck as it is unable to find a safe $\theta_+^\star$. This is especially relevant at the beginning of the simulation in case $\mathcal{D}_0$ is empty, i.e., data is scarce and the GP approximation is poor. Currently, one can mitigate this issue by backtracking (i.e., iteratively reducing) the requested safety probability $\beta$ to smaller and smaller values till feasibility is recovered, as shown in lines \ref{line:algorithm:backtracking_startline}-\ref{line:algorithm:backtracking_endline}. Of course, this trades off safety with feasibility, which might be undesirable in cases where the former must be guaranteed with high enough certainty. However, one could easily envision methods for alleviating this issue by, e.g., injecting prior knowledge via the GP prior mean or the initial dataset $\mathcal{D}_0$, or, alternatively, assuming the initial condition $\theta_0$ to be safe with unitary probability (thus acting as a sort of backup configuration).
\end{rem}

\begin{rem}
    While GPs are convenient because they allow a deterministic formulation of \eqref{eq:algorithm:constrainted_update:gp_constraints} that naturally takes estimation uncertainty into account, they are not per se a mandatory component of the proposed approach. One could indeed envision the use of different regressor types with different advantages and disadvantages, such as, to mention few, Inverse Distance Weighting interpolants \citep{zhu_cglisp_2022}, whose output in range $[0, 1]$ can be directly interpreted as probability, Support Vector Machines \citep{smola_tutorial_2004}, that perform well with small/medium-size training sets but do not quantify uncertainty, or Bayesian Neural Networks \citep{jospin_handson_2022}, which are powerful tools in approximating complex distributions but require much larger amounts of training data. Moreover, different types of Gaussian Process regression are available that address the weaknesses of vanilla GPs that may emerge during training, such as the sparse \citep{csato_sparse_2002} and warped variants \citep{snelson_warped_2003}, which are suitable for training on larger datasets or allow for a non-Gaussian process by warping variables to a latent space, respectively.
\end{rem}

\begin{rem}
    While update \eqref{eq:algorithm:constrainted_update} is performed at every iteration of Algorithm \ref{algo:algorithm:main}, it can be in principle performed at lower frequencies. This is especially relevant in those RL tasks in which convergence is difficult to achieve and can benefit from the employment of experience replay buffers to stabilise learning.
\end{rem}

\section{Numerical Experiment} \label{section:experiment}
To put our proposed algorithm to the test, it was implemented and simulated in a numerical experiment of a quadrotor drone application adapted from \cite{wabersich_cautious_2022}. In the following sections, we detail the system, the controller and the safe learning algorithm, and present the final results.

\subsection{System Description}
The aim of the task is for the drone to reach a final destination in the fastest and most accurate way, while avoiding violations of safe-critical boxing constraints on the states and control actions. Wind disturbances at different altitudes influence the drone, which add nonlinear effects to the dynamics. The goal of the RL agent is then double: on one side, it tunes the parameters of the MPC controller to achieve stable flight and better performance in targeting the final position; on the other side, it learns to avoid constraint violations as much as possible, despite wind disturbances and while still accomplishing the task.
\par
The discrete-time dynamics of the drone are of the form
\begin{equation}
    x_{t+1} = A x_t + B u_t + C \Psi(x_t) w_t + G,
\end{equation}
where the state $x \in \mathbb{R}^{10}$ contains the pose (position, velocity, attitude, and its rate) of the quadrotor, the control action $u \in \mathbb{R}^3$ dictates the desired pitch/roll attitude and vertical acceleration, and system matrices $A$ and $B$ describe the dynamics of the quadrotor around the hovering state, and vector $G$ represents the constant gravity pull. See Appendix \ref{section:appendix:qudrotor_dynamics} and \cite{bouffard_onboard_2012} for a more detailed insight on the model. The additional term $C \Psi(x_t)$ models nonlinear wind disturbances at time step $t$: $\Psi(x) \in \mathbb{R}^{3}$ is a state-dependent vector simulating three different wind currents at different altitudes, each implemented as a squared exponential basis function with unknown hyperparameters, whereas matrix $C$ modulates the influence of each gust on each state, and $w_t \sim \mathcal{U}(0, 1)$ acts as a source of randomness. Box constraints $h$ are imposed on the position, pitch, roll, as well as on the three control actions. The RL stage cost takes into account, at each step, the error from the target position $s_\text{f}$, the control action usage, as well as constraint violations
\begin{equation}
    L(s,a) = \norm[2]{s - s_\text{f}}^2 + c_1 \norm[2]{a}^2 + c_2^\top \max{\bigl(0, h\left(s,a\right)\bigr)},
\end{equation}
where scalar $c_1$ and vector $c_2$ weigh the different contributions.

\subsection{MPC Function Approximation}
The controller chosen for this task consists of an MPC scheme based on \eqref{eq:background:V:scheme}: the stage/final costs are tracking costs (in both states and actions) but their cost matrices are fixed/non-learnable, and no initial cost is used, i.e., $\lambda_\theta(x_0) = 0$; the parametric prediction model is $f_\theta(x,u) = A_\theta x + B_\theta u + G_\theta$; the constraints are parametrized by means of a backoff parameter $h_\theta(x,u) = (1 + \delta) h(x,u)$ and are made soft in the MPC to avoid infeasibilities. Eight different system parameters in $A_\theta$, $B_\theta$ and $G_\theta$ affect the dynamics of the quadrotor drone but, out of these, the learning is restricted exclusively to the gravitational pull constant $g$ and the vertical thruster coefficient $K_z$. Additionally, the constraint backoff $\delta$ is made learnable, i.e., $\theta = \begin{bmatrix} g & K_z & \delta \end{bmatrix}^\top$, while all the other parameters in the drone model and optimization are kept fixed during learning. To better test the viability of the proposed approach, all eight dynamics parameters in the MPC scheme are initialized wrongly (with an error $\approx \pm 20\%$), so as to lead to a controller with initial poor performance and constraint violations. Therefore, the agent has to achieve better performance and safety by only tuning $\theta$, a small subset of all the dynamics parameters. It is worth highlighting the notion that this kind of parameter tuning differs in essence from the classical system identification, where parameters are chosen in such a way as to match the system's behaviour with observed data (e.g., in the sense of the prediction mean-squared-error), and a controller is only subsequently designed, in a possibly iterative procedure. On the contrary, here the adjustments of the learnable parameters are directly carried out with the goal of optimizing performance, regardless of the model's prediction accuracy. This approach is sometimes called Identification for Control, according to which the best model for control needs not be the one providing the best predictions, but the best performance on the true system \citep{piga_performance_2019}. The two identification procedures are not mutually exclusive and can be appropriately incorporated in a unique approach, as done in \cite{martinsen_combining_2020}.

\subsection{Safe RL Algorithm}
For learning safety constraints \eqref{eq:background:max_safety_constraints}, one independent GP is employed to model each constraint. A zero prior mean function was assumed, whereas the kernel function was selected as the sum of the squared exponential kernel and a white noise kernel:
\begin{equation}
    \nu(\theta, \theta^\ast) = \sigma_1^2 \exp{\left(-\frac{1}{2 \ell^2}\norm[2]{\theta - \theta^\ast}^2\right)} +  \sigma_2^2 I_{n_\theta} \delta_{\theta \theta^\ast},
\end{equation}
where $\ell$ is the length scale and $\sigma_1$ is the multiplier of the exponential, $\sigma_2$ denotes the white noise level, and $\delta_{\theta \theta^\ast}$ is the Kronecker delta. As per Algorithm \ref{algo:algorithm:main}, the GP regressors are trained on past observed data before each update of the MPC parameters. For the RL updates, a second-order LSTD Q-learning algorithm \citep{lagoudakis_leastsquares_2002} is employed to find the parameter vector $\theta$ that minimizes closed-loop performance under the policy $\pi_\theta(s)$. Q-learning is a classical RL algorithm that tries to find the parametrization which best fits the action-value function to the observed data, thus indirectly finding the optimal policy, and has shown promising results also with MPC \citep{esfahani_approximate_2021}. The second-order Newton's method ensures faster convergence, for which gradient $p$ and approximate Hessian $H$ are found as
\begin{align}
    \delta_i &= L(s_i,a_i) + \gamma V_\theta(s_{i+1}) - Q_\theta(s_i,a_i) \\
    p &= - \sum_{i=1}^{m}{\delta_i \nabla_\theta Q_\theta(s_i,a_i)} \\
    H &= \sum_{i=1}^{m}{
        \nabla_\theta Q_\theta(s_i,a_i) \nabla_\theta Q_\theta^\top(s_i,a_i) -
        \delta_i \nabla^2_\theta Q_\theta(s_i,a_i)
    },
\end{align}
where $H$ is modified when needed to be positive definite \cite[Algorithm~3.3]{nocedal_numerical_2006}, and the gradient $\nabla_\theta Q_\theta$ is computed via nonlinear programming sensitivity analysis \citep{buskens_sensitivity_2001}. As in \cite{zanon_safe_2021}, exploratory behaviour is ensured during learning by adding to the MPC objective \eqref{eq:background:V:obj} the perturbation term $q^\top u_0$, with $q$ randomly chosen. Lastly, in the constrained update \eqref{eq:algorithm:constrainted_update}, the target safety level $\beta$ is set to $0.9$, but backtracking on this level is enabled in case of infeasibility.

\subsection{Results}
The experiment was implemented in Python leveraging the symbolic framework CasADi \citep{andersson_casadi_2019} and its interface to the IPOPT solver \citep{wachter_implementation_2006}. The source code and simulation results can be found in the following repository: \texttt{https://github.com/} \texttt{FilippoAiraldi/learning-safety-in-mpc-based-rl}.

To investigate the effectiveness of the proposed algorithm, simulations were carried out for the LSTD Q-learning algorithm both with and without our GP-based safety enhancement. The safe variant of the algorithm was also tested with zero initial knowledge (i.e., $\mathcal{D}_0$ empty), and with limited prior knowledge in the form of 5 datapoints picked randomly from previous simulations. Where relevant, we also compare the algorithm with a baseline, in which the MPC controller is initialized with perfect knowledge on the system model. Each algorithm is averaged over 100 different simulations for 50 learning episodes. In the figures, average results are plotted with one standard deviation.

\begin{figure}
    \centering
    \scalebox{1}{\input{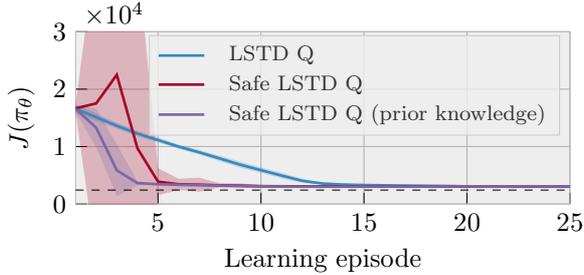}}
    \caption{Comparison in performance between the original LSTD Q-learning algorithm and its GP-based safe variant (with and without prior knowledge). In dashed black, the baseline cumulative cost when exact knowledge of the system is given to the MPC controller.}
    \label{fig:performance}
\end{figure}
As shown in Figure \ref{fig:performance}, despite starting from an unfavourable initial configuration $\theta$ that yields large constraint violations, both safe and unsafe algorithms are able to recover and converge to a performance that is just $\approx 20\%$ worse than the baseline. Note that this recovery is achieved through adjusting only a limited subset of the system dynamics parameters. However, the convergence speed of the two substantially differs, the safe variants being much faster. This is explained by the fact that already in the first few iterations the safety mechanism is able to approximate the safe set $\mathcal{S}_{\mathcal{D}}$, even if only locally, and to steer the learning towards parameters $\theta$ that are likely to yield safer policies. However, we note that at the very beginning of the learning process the zero-knowledge safe algorithm suffers from much higher variance in the results compared to its counterpart initialized with some prior information. This confirms that at the beginning the ability to predict safety is extremely poor and can even damage performance.

\begin{figure}
    \centering
    \scalebox{1}{\input{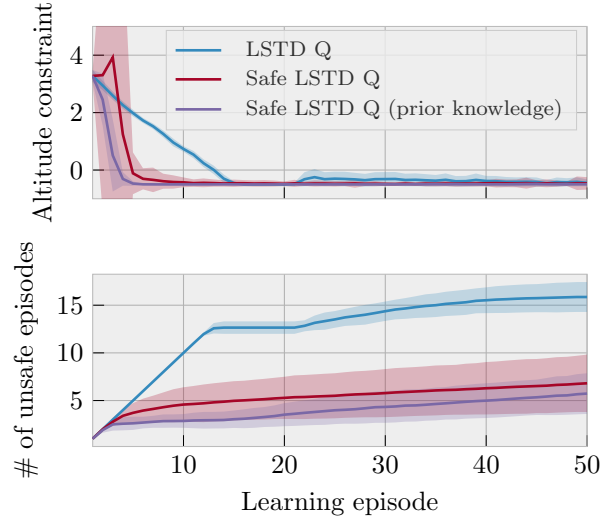}}
    \caption{Comparison in (top) violations of the altitude constraint, where positive values imply violation, and (bottom) the cumulative number of unsafe episodes.}
    \label{fig:safety}
\end{figure}
Figure \ref{fig:safety} shows the effectiveness of our method in encouraging safety. At the end of the learning process, we witnessed a $57\%$ and $64\%$ reduction in the total number of unsafe episodes (episodes in which constraint violation occurs) for the two safe algorithms compared to the unsafe LSTD Q-learning variant. These results are achieved in a purely data-driven fashion and with no or little previous information on safety. Furthermore, constraint violations are recovered much faster, especially in the quadrotor altitude, which is the state variable most prone to violations. This explains why the performance of the safe algorithms shows faster convergence to better suboptimal policies.

\begin{figure}
    \centering
    \scalebox{1}{\input{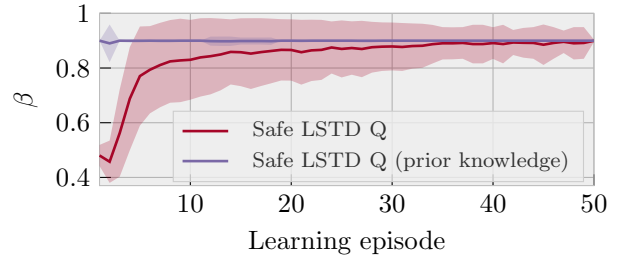}}
    \caption{Backtracked safety probability $\beta$ during learning.}
    \label{fig:backtracking}
\end{figure}
Lastly, Figure \ref{fig:backtracking} shows the impact of the backtracking on $\beta$, as explained in Remark \ref{remark:algorithm:backtracking_beta}. In particular, it shows on average during learning at which probability Algorithm \ref{algo:algorithm:main} had to decrease $\beta$ in order to obtain the feasibility of the constrained update \eqref{eq:algorithm:constrainted_update}. In the context of zero initial information, at the beginning $\beta$ has to be substantially decreased to achieve feasible updates, while at convergence, when enough data has been collected to properly train the GPs, the required probability of $0.9$ is achieved. Conversely, the simulations with initial knowledge seems to confirm that providing a small $\mathcal{D}_0$ to pre-train the GP-based safety constraints \eqref{eq:algorithm:constrainted_update:gp_constraints} is an effective way to avoid backtracking on $\beta$ to recover feasible updates.

\section{Conclusions} \label{section:conclusions}
In this paper, we propose a method that enables safety in the context of MPC-based RL via GP regression. It learns and encourages safety exclusively from observed past trajectories, without the need of, e.g., a priori information on the system or estimation of disturbances affecting its dynamics. The algorithm is straightforward and can be implemented without extensive customization on top of several RL agents, such as Q-learning and policy gradient methods. The numerical experiment for the quadrotor control application demonstrated that the safety mechanism successfully reduces the number of unsafe episodes during learning and also promotes faster convergence by avoiding unnecessary constraint violations.

Nonetheless, there are still open points regarding the proposed method. The necessity of backtracking $\beta$ is somewhat undesirable, since it deteriorates safety. To avoid that, simulations where initial knowledge on safety is provided showed how this can alleviate the need for reducing $\beta$. However, even so, the algorithm may still get stuck and become infeasible for any $\beta$. An open question is how to avoid such situations. Additionally, it would be desirable to integrate this algorithm with more complex MPC schemes (e.g., robust or stochastic) where also other components (e.g., the disturbance set) may be learnt online in a probabilistic way. Finally, readers acquainted with Bayesian Optimization may have noticed a natural similarity between Algorithm \ref{algo:algorithm:main} and its sampling-based counterparts, which we took inspiration from. An interesting point would be therefore to compare the efficacy, both in terms of performance and safety, of our gradient-based algorithm with safe BO algorithms.

\begin{ack}
    This research is part of a project that has received funding from the European Research Council (ERC) under the European Union’s Horizon 2020 research and innovation programme (Grant agreement No. 101018826 - CLariNet).
\end{ack}

\bibliography{references}

\begin{thebibliography}{30}
\providecommand{\natexlab}[1]{#1}
\providecommand{\url}[1]{\texttt{#1}}
\providecommand{\urlprefix}{URL }
\expandafter\ifx\csname urlstyle\endcsname\relax
  \providecommand{\doi}[1]{doi:\discretionary{}{}{}#1}\else
  \providecommand{\doi}{doi:\discretionary{}{}{}\begingroup
  \urlstyle{rm}\Url}\fi

\bibitem[{Andersson et~al.(2019)Andersson, Gillis, Horn, Rawlings, and
  Diehl}]{andersson_casadi_2019}
Andersson, J.A.E., Gillis, J., Horn, G., Rawlings, J.B., and Diehl, M. (2019).
\newblock {CasADi}: a software framework for nonlinear optimization and optimal
  control.
\newblock \emph{Mathematical Programming Computation}, 11(1), 1--36.

\bibitem[{Bouffard(2012)}]{bouffard_onboard_2012}
Bouffard, P. (2012).
\newblock \emph{On-board Model Predictive Control of a Quadrotor Helicopter:
  Design, Implementation, and Experiments}.
\newblock Master's thesis, EECS Department, University of California, Berkeley.

\bibitem[{Brunke et~al.(2022)Brunke, Greeff, Hall, Yuan, Zhou, Panerati, and
  Schoellig}]{brunke_safe_2022}
Brunke, L., Greeff, M., Hall, A.W., Yuan, Z., Zhou, S., Panerati, J., and
  Schoellig, A.P. (2022).
\newblock Safe learning in robotics: from learning-based control to safe
  reinforcement learning.
\newblock \emph{Annual Review of Control, Robotics, and Autonomous Systems},
  5(1), 411--444.

\bibitem[{B{\"u}skens and Maurer(2001)}]{buskens_sensitivity_2001}
B{\"u}skens, C. and Maurer, H. (2001).
\newblock Sensitivity analysis and real-time optimization of parametric
  nonlinear programming problems.
\newblock In M.~Gr{\"o}tschel, S.O. Krumke, and J.~Rambau (eds.), \emph{Online
  Optimization of Large Scale Systems}, 3--16. Springer, Berlin, Heidelberg.

\bibitem[{Busoniu et~al.(2017)Busoniu, Babuska, De~Schutter, and
  Ernst}]{busoniu_reinforcement_2017}
Busoniu, L., Babuska, R., De~Schutter, B., and Ernst, D. (2017).
\newblock \emph{Reinforcement learning and dynamic programming using function
  approximators}.
\newblock CRC press.

\bibitem[{Csató and Opper(2002)}]{csato_sparse_2002}
Csató, L. and Opper, M. (2002).
\newblock Sparse on-line gaussian processes.
\newblock \emph{Neural Computation}, 14(3), 641--668.

\bibitem[{Esfahani et~al.(2021)Esfahani, Kordabad, and
  Gros}]{esfahani_approximate_2021}
Esfahani, H.N., Kordabad, A.B., and Gros, S. (2021).
\newblock Approximate robust {NMPC} using reinforcement learning.
\newblock In \emph{2021 European Control Conference (ECC)}, 132--137.

\bibitem[{Gros and Zanon(2022)}]{gros_learning_2022}
Gros, S. and Zanon, M. (2022).
\newblock Learning for {MPC} with stability \& safety guarantees.
\newblock \emph{Automatica}, 146, 110598.

\bibitem[{Gros and Zanon(2020)}]{gros_datadriven_2020}
Gros, S. and Zanon, M. (2020).
\newblock Data-driven economic {NMPC} using reinforcement learning.
\newblock \emph{IEEE Transactions on Automatic Control}, 65(2), 636--648.

\bibitem[{Hewing et~al.(2020{\natexlab{a}})Hewing, Kabzan, and
  Zeilinger}]{hewing_cautious_2020}
Hewing, L., Kabzan, J., and Zeilinger, M.N. (2020{\natexlab{a}}).
\newblock Cautious model predictive control using gaussian process regression.
\newblock \emph{IEEE Transactions on Control Systems Technology}, 28(6),
  2736--2743.

\bibitem[{Hewing et~al.(2020{\natexlab{b}})Hewing, Wabersich, Menner, and
  Zeilinger}]{hewing_learning_2020}
Hewing, L., Wabersich, K.P., Menner, M., and Zeilinger, M.N.
  (2020{\natexlab{b}}).
\newblock Learning-based model predictive control: toward safe learning in
  control.
\newblock \emph{Annual Review of Control, Robotics, and Autonomous Systems},
  3(1), 269--296.

\bibitem[{Jospin et~al.(2022)Jospin, Laga, Boussaid, Buntine, and
  Bennamoun}]{jospin_handson_2022}
Jospin, L.V., Laga, H., Boussaid, F., Buntine, W., and Bennamoun, M. (2022).
\newblock Hands-on bayesian neural networks—a tutorial for deep learning
  users.
\newblock \emph{IEEE Computational Intelligence Magazine}, 17(2), 29--48.
\newblock \doi{10.1109/MCI.2022.3155327}.

\bibitem[{Koller et~al.(2018)Koller, Berkenkamp, Turchetta, and
  Krause}]{koller_learning_2018}
Koller, T., Berkenkamp, F., Turchetta, M., and Krause, A. (2018).
\newblock Learning-based model predictive control for safe exploration.
\newblock In \emph{2018 IEEE Conference on Decision and Control (CDC)},
  6059--6066.

\bibitem[{Krishnamoorthy and Doyle(2022)}]{krishnamoorthy_safe_2022}
Krishnamoorthy, D. and Doyle, F.J. (2022).
\newblock Safe bayesian optimization using interior-point methods—applied to
  personalized insulin dose guidance.
\newblock \emph{IEEE Control Systems Letters}, 6, 2834--2839.

\bibitem[{Lagoudakis et~al.(2002)Lagoudakis, Parr, and
  Littman}]{lagoudakis_leastsquares_2002}
Lagoudakis, M.G., Parr, R., and Littman, M.L. (2002).
\newblock Least-squares methods in reinforcement learning for control.
\newblock In \emph{Methods and Applications of Artificial Intelligence},
  249--260. Springer Berlin Heidelberg, Berlin, Heidelberg.

\bibitem[{Martinsen et~al.(2020)Martinsen, Lekkas, and
  Gros}]{martinsen_combining_2020}
Martinsen, A.B., Lekkas, A.M., and Gros, S. (2020).
\newblock Combining system identification with reinforcement learning-based
  {MPC}.
\newblock \emph{IFAC-PapersOnLine}, 53(2), 8130--8135.
\newblock 21st IFAC World Congress.

\bibitem[{Mesbah et~al.(2022)Mesbah, Wabersich, Schoellig, Zeilinger, Lucia,
  Badgwell, and Paulson}]{mesbah_2022_fusion}
Mesbah, A., Wabersich, K.P., Schoellig, A.P., Zeilinger, M.N., Lucia, S.,
  Badgwell, T.A., and Paulson, J.A. (2022).
\newblock Fusion of machine learning and {MPC} under uncertainty: What advances
  are on the horizon?
\newblock In \emph{2022 American Control Conference (ACC)}, 342--357.

\bibitem[{Nocedal and Wright(2006)}]{nocedal_numerical_2006}
Nocedal, J. and Wright, S.J. (2006).
\newblock \emph{Numerical Optimization}.
\newblock Springer, New York, NY, USA, 2nd edition.

\bibitem[{Piga et~al.(2019)Piga, Forgione, Formentin, and
  Bemporad}]{piga_performance_2019}
Piga, D., Forgione, M., Formentin, S., and Bemporad, A. (2019).
\newblock Performance-oriented model learning for data-driven {MPC} design.
\newblock \emph{IEEE Control Systems Letters}, 3(3), 577--582.

\bibitem[{Rasmussen and Williams(2005)}]{rasmussen_gaussian_2005}
Rasmussen, C.E. and Williams, C.K.I. (2005).
\newblock \emph{Gaussian Processes for Machine Learning}.
\newblock MIT Press.

\bibitem[{Schreiter et~al.(2015)Schreiter, Nguyen-Tuong, Eberts, Bischoff,
  Markert, and Toussaint}]{schreiter_safe_2015}
Schreiter, J., Nguyen-Tuong, D., Eberts, M., Bischoff, B., Markert, H., and
  Toussaint, M. (2015).
\newblock Safe exploration for active learning with gaussian processes.
\newblock In \emph{Machine Learning and Knowledge Discovery in Databases},
  133--149. Springer International Publishing.

\bibitem[{Smola and Sch{\"o}lkopf(2004)}]{smola_tutorial_2004}
Smola, A.J. and Sch{\"o}lkopf, B. (2004).
\newblock A tutorial on support vector regression.
\newblock \emph{Statistics and Computing}, 14(3), 199--222.

\bibitem[{Snelson et~al.(2003)Snelson, Ghahramani, and
  Rasmussen}]{snelson_warped_2003}
Snelson, E., Ghahramani, Z., and Rasmussen, C. (2003).
\newblock Warped gaussian processes.
\newblock In \emph{Advances in Neural Information Processing Systems},
  volume~16. MIT Press.

\bibitem[{Sorourifar et~al.(2021)Sorourifar, Makrygirgos, Mesbah, and
  Paulson}]{sorourifar_data_2021}
Sorourifar, F., Makrygirgos, G., Mesbah, A., and Paulson, J.A. (2021).
\newblock A data-driven automatic tuning method for {MPC} under uncertainty
  using constrained bayesian optimization.
\newblock \emph{IFAC-PapersOnLine}, 54(3), 243--250.
\newblock 16th IFAC Symposium on Advanced Control of Chemical Processes ADCHEM
  2021.

\bibitem[{Sutton and Barto(2018)}]{sutton_reinforcement_2018}
Sutton, R.S. and Barto, A.G. (2018).
\newblock \emph{Reinforcement Learning: An Introduction}.
\newblock MIT Press, Cambridge, MA, USA.

\bibitem[{Turchetta et~al.(2019)Turchetta, Berkenkamp, and
  Krause}]{turchetta_safe_2019}
Turchetta, M., Berkenkamp, F., and Krause, A. (2019).
\newblock Safe exploration for interactive machine learning.
\newblock In \emph{Advances in Neural Information Processing Systems},
  volume~32. Curran Associates, Inc.

\bibitem[{Wabersich and Zeilinger(2022)}]{wabersich_cautious_2022}
Wabersich, K.P. and Zeilinger, M.N. (2022).
\newblock Cautious bayesian {MPC}: regret analysis and bounds on the number of
  unsafe learning episodes.
\newblock \emph{IEEE Transactions on Automatic Control}, 1--8.

\bibitem[{W{\"a}chter and Biegler(2006)}]{wachter_implementation_2006}
W{\"a}chter, A. and Biegler, L.T. (2006).
\newblock On the implementation of an interior-point filter line-search
  algorithm for large-scale nonlinear programming.
\newblock \emph{Mathematical Programming}, 106(1), 25--57.

\bibitem[{Zanon and Gros(2021)}]{zanon_safe_2021}
Zanon, M. and Gros, S. (2021).
\newblock Safe reinforcement learning using robust {MPC}.
\newblock \emph{IEEE Transactions on Automatic Control}, 66(8), 3638--3652.

\bibitem[{Zhu et~al.(2022)Zhu, Piga, and Bemporad}]{zhu_cglisp_2022}
Zhu, M., Piga, D., and Bemporad, A. (2022).
\newblock {C-GLISp}: preference-based global optimization under unknown
  constraints with applications to controller calibration.
\newblock \emph{IEEE Transactions on Control Systems Technology}, 30(5),
  2176--2187.

\end{thebibliography}

\appendix
\section{Quadrotor's System Dynamics} \label{section:appendix:qudrotor_dynamics}
As described in Section \ref{section:experiment}, the discrete-time dynamics of the quadrotor drone are
\begin{equation*}
    x_{t+1} = A x_t + B u_t + C \Psi(x_t) w_t + G,
\end{equation*}
where the state consists of the 3D positions and velocities, as well as roll and pitch attitudes and rates, i.e.,
\begin{equation*}
    x = \begin{bmatrix}
        p_x & p_y & p_z & v_x & v_y & v_z & a_p & a_r & r_p & r_r
    \end{bmatrix}^\top,
\end{equation*}
and the control action is desired pitch and roll and vertical acceleration, i.e.,
\begin{equation*}
    u = \begin{bmatrix}
        u_p & u_r & u_z
    \end{bmatrix}^\top.
\end{equation*}
In particular, the dynamics can be described with the following matrices:
\begin{align*}
    A &= I_{10} + T_\text{s} \begin{bmatrix}
        0 & 0 & 0 & 1 & 0 & 0 & 0            & 0            & 0 & 0 \\
        0 & 0 & 0 & 0 & 1 & 0 & 0            & 0            & 0 & 0 \\
        0 & 0 & 0 & 0 & 0 & 1 & 0            & 0            & 0 & 0 \\
        0 & 0 & 0 & 0 & 0 & 0 & g            & 0            & 0 & 0 \\
        0 & 0 & 0 & 0 & 0 & 0 & 0            & g            & 0 & 0 \\
        0 & 0 & 0 & 0 & 0 & 0 & 0            & 0            & 0 & 0 \\
        0 & 0 & 0 & 0 & 0 & 0 & -d_p         & 0            & 1 & 0 \\
        0 & 0 & 0 & 0 & 0 & 0 & 0            & -d_r         & 0 & 1 \\
        0 & 0 & 0 & 0 & 0 & 0 & -d_{\dot{p}} & 0            & 0 & 0 \\
        0 & 0 & 0 & 0 & 0 & 0 & 0            & -d_{\dot{r}} & 0 & 0
    \end{bmatrix} \\
    B &= T_\text{s} \begin{bmatrix}
            & 0_{5 \times 3} &     \\
        0   & 0              & K_z \\
            & 0_{2 \times 3} &     \\
        K_p & 0              & 0   \\
        0   & K_r            & 0
    \end{bmatrix} \\
    G &= - T_\text{s} \begin{bmatrix}
        0_{1 \times 5} & g & 0_{1 \times 4}
    \end{bmatrix}^\top,
\end{align*}
where the various parameters can be found in table \ref{tb:appendix:quadrotor_parameters}.
\begin{table}
    \begin{center}
        \caption{Quadrotor's Parameters} \label{tb:appendix:quadrotor_parameters}
        \begin{tabular}{cc}
            Parameter                        & Symbol \\ \hline
            sampling time                    & $T_\text{s}$ \\
            gravitational constant           & $g$ \\
            pitch/roll attitude coefficients & $d_{\{p, r\}}$ \\
            pitch/roll rate coefficients     & $d_{\{\dot{p}, \dot{r}\}}$ \\
            control gains                    & $K_{\{p, r, z\}}$ \\ \hline
        \end{tabular}
    \end{center}
\end{table}
The nonlinear term $C \Psi(x)$ is used to model three wind currents disturbing the dynamics:
\begin{align*}
    C &= T_\text{s} \begin{bmatrix}
        c_{11} & c_{12}         & c_{13} \\
        c_{21} & c_{22}         & c_{23} \\
        c_{31} & c_{32}         & c_{33} \\
               & 0_{3 \times 3} &        \\
        c_{71} & c_{72}         & c_{73} \\
        c_{81} & c_{82}         & c_{83} \\
               & 0_{2 \times 3} &
    \end{bmatrix} \\
    \Psi(x) &= \begin{bmatrix}
        \psi_1(p_z) & \psi_2(p_z) & \psi_3(p_z)
    \end{bmatrix}^\top,
\end{align*}
where $\psi_i(p_z) = \exp{\left(-(p_z - b_i)^2\right)}$ represents the effect of the $i$-th gust at altitude $b_i$ given the current vertical position $p_z$ of the drone.

\end{document}